\documentclass[12pt]{article}
\usepackage{epsfig,picinpar,floatflt,amssymb}
\def\to{\rightarrow}

\def\bd{\begin{displaystyle}}
\def\ed{\end{displaystyle}}
\def\EQ{\begin{equation}}
\def\EN{\end{equation}}
\def\bea{\begin{eqnarray}}
\def\eea{\end{eqnarray}}

\def\hs{\hspace{0.1in}}

\begin{document}
\oddsidemargin 5mm
\setcounter{page}{0}
\newpage     
\setcounter{page}{0}
\begin{titlepage}
\begin{flushright}
ISAS/EP/97/153 \\
IC/97/199
\end{flushright}
\vspace{0.5cm}
\begin{center}
{\large {\bf The Quantum Mechanical Potential \\
for the Prime Numbers}}\\
\vspace{1.5cm}
{\bf G. Mussardo$^{a,b,c}$} \\
\vspace{0.8cm}
$^a${\em International School for Advanced Studies, Via Beirut 2-4, 
34013 Trieste, Italy} \\ 
$^b${\em Istituto Nazionale di Fisica Nucleare, Sezione di Trieste}\\
$^c${\em The Abdus Salam International Centre for Theoretical Physics \\
Strada Costiera 12, 34014 Trieste }\\
\end{center}
\vspace{6mm}
\begin{abstract}
\noindent
A simple criterion is derived in order that a number sequence ${\cal
S}_n$ is a permitted spectrum of a quantized system. The sequence of
the prime numbers fulfils the criterion and the corresponding one-dimensional 
quantum potential is explicitly computed in a semi-classical approximation.
The existence of such a potential implies that the primality testing 
can in principle be resolved by the sole use of physical laws.
\end{abstract}
\vspace{5mm}
\end{titlepage}
\newpage

\setcounter{footnote}{0}
\renewcommand{\thefootnote}{\arabic{footnote}}


In the realm of arithmetic, or rather number theory, one often
encounters interesting number sequences ${\cal S}_n$ (see, for
instance \cite{Rose,Ore,Schroeder}). For many of the sequences, it is 
their recreational aspect which renders them interesting, as in the 
case of the polygonal numbers $p_s^n$ \cite{polygonal}. For other sequences, 
instead, their interesting aspects consist in the influence which 
they still exert over devotees of number theory as sources of delighful 
surprises or intriguing properties, such as the perfect number sequence
\cite{perfect} 
\EQ
{\cal P}_n = 1, 6, 28, 496, 8128, \ldots 
\label{perfectnumbers}
\EN
or Fibonacci numbers \cite{Fibonacci} 
\EQ
F_n = 1, 1, 2, 3, 5, 8, 13, 21, \ldots
\label{Fibonaccinumbers}
\EN 
Finally, it is probably fair to say that the most important sequence of
integers in number theory -- and still one of the most elusive -- is
represented by the class of primes 
\EQ
P_n = 2, 3, 5, 7, 11, 13, 17, 19, 23, 29, \ldots 
\label{primenumbers}
\EN 
No closed formula is known for the $n$-th prime number. However, by 
accepting an approximation of their value within a $10\%$ margin, it is 
known that the $n$-th prime number may be expressed in terms of the simple
formula   
\EQ
P_n \simeq \,\,n \,\log n \,\,\,.
\label{primi}
\EN

In the realm of physics, or rather quantum mechanics, it is also 
very natural to come upon discrete sequences of numbers. For instance, 
these may be the allowed energy values of a given system, once its 
quantization is carried out. This being so, it is then natural to wonder
whether it would always be possible to design a physical device in
such a way that the resulting quantized energy levels (in appropriate 
units) concide with an assigned sequence of numbers taken from the 
arithmetic realm\footnote{By physical device we intend any apparatus 
(made by electric field traps, etc.) which can be parameterized in terms of a 
potential $V(x)$ entering the Schr\"{o}dinger equation. For simplicity
we consider only the one-dimensional case in this paper. In the following, 
the energy levels will often be denoted by their corresponding number
sequences since the correct physical dimensions can always be restored by 
appropriate dimensional factors.}. 

The answer to the above question is actually negative, i.e. {\em not all} 
number sequences ${\cal S}_n$ can play the role of discrete energy
eigenvalues of a quantized system. The easiest way to show this result is
to derive a bound on the growing rate of energy eigenvalues of
one--dimensional quantum hamiltonian by employing the semi--classical
quantization formula\footnote{To simplify the following expressions we 
will assume $V(x) = V(-x)$. The constant $\alpha$ in eq.(\ref{quantization}) 
will not play any significative role in the following and can be taken 
equal to $0$.}
\EQ
\oint p(x) dx = (n + \alpha) h \,\, , 
\label{quantization}
\EN 
where $p(x) = \sqrt{2 m \left[E - V(x)\right]}$. Let us initially
consider the class of potentials 
\EQ
V_r(x) \,=\, \lambda \mid x\mid^r \,\,\,.
\label{potential}
\EN  
Substituting (\ref{potential}) into eq.\,(\ref{quantization}), it is easy 
to obtain that the energy levels associated to these potentials scale as    
\EQ
E_n \,=\,A_r\, 
\,n^{\frac{2 r}{r+2}} \,\,\, ,
\label{scale} 
\EN 
where 
\EQ
A_r \,=\, 
\left[\frac{\lambda^{1/r} h \Gamma\left(\frac{1}{r} +
\frac{3}{2}\right)}  {2 \sqrt{2 m \pi} \Gamma\left(\frac{1}{r} +1\right)}
\right]^{\frac{2r}{r+2}} \,\,\,.
\label{constant}
\EN 
From (\ref{scale}) we see that the quantized energies relative to 
such potentials cannot grow faster than $E_n \sim n^2$, i.e. they 
always satisfy the bound 
\EQ
E_n \,\leq \, C\, n^2 \,\,\,.
\label{bound}
\EN
The faster the variation of the potential, the faster is of the
increase of the energy eigenvalues and the bound is therefore saturated 
for the quickest increasing function of the family (\ref{potential}), i.e. 
the potential well 
\EQ
V(x) \, = \, \lim_{r\to\infty} V_r(x) \,=\,
\left\{
\begin{array}{ll}
0 \, , & \mbox {if $|x| < 1$} \\
\infty \, ,& \mbox {otherwise}
\end{array}
\right.
\EN 

It is easy to show that the validity of the bound (\ref{bound}) 
extends beyond the polynomial potentials which we used for its 
derivation\footnote{It is simple to see, for instance, that even for 
a potential which increases exponentially $V(x) \sim \mu \exp(\beta |x|)$ 
for $|x| \to \infty$, the eigenvalues cannot grow more than quadratically.}. 
In fact, it can be easily argued that the bound (\ref{bound}) is actually
a general feature of all discrete spectra of one-dimensional quantum 
hamiltonian.  

It should be clear at this stage that in order for a number sequence
${\cal S}_n$ to be a permitted spectrum of a quantized system, it must 
increase at a rate slower than $n^2$. This implies, for instance, that
a one--dimensional quantum mechanical system that has Fibonacci numbers 
(\ref{Fibonaccinumbers}) as energy eigenvalues cannot exist. By the same 
token, other exponentially increasing number sequences are ruled out as well. 
Viceversa, quantum mechanical systems which allow, for instance, the 
polygonal numbers (\ref{perfectnumbers}) as possible energy eigenvalues
must exist. The same is also true for the sequence of the prime numbers, 
since their behaviour (\ref{primi}) is sufficiently mild so as not to 
violate the bound (\ref{bound}). 

How can the potential corresponding to an allowed sequence of numbers be 
explicitly computed? As it is well known, the answer can be given in the 
context of semi-classical quantization, i.e. by inverting 
eq.\,(\ref{quantization}) 
\EQ
x(V) \,=\,
\frac{\hbar}{\sqrt{2 m}} 
\int_{E_0}^V \frac{dE}{\frac{dE}{dn} 
\sqrt{V - E}} \,\,\, ,
\label{inverse}
\EN
where $E_0$ sets the zero of the energy scale. The potential $V(x)$ 
is thus finally obtained in terms of the inverse function of $x(V)$. 

Let us now apply the above semi-classical formula (\ref{inverse}) in order 
to find the quantum mechanical potential which has the prime numbers $P_n$ as 
its energy eigenvalues\footnote{To restore the correct dimension, 
it may be convenient to introduce -- in perfect analogy with the harmonic 
oscillator -- a frequency $\omega$ and express the energy eigenvalues as 
$E_n = \hbar \omega \,P_n$.} $E_n$. Let us denote this potential as
$W(x)$. As mentioned before, a closed formula for the $n$-th prime number
is not known but this is not an obstacle to obtain $W(x)$. In
fact, there is a great deal of information about $\pi(N)$, i.e. the function
which counts the prime numbers smaller or equal to a given number $N$. 
Since the formula of the $n$-th prime number is nothing but the inverse 
of the function $\pi(N)$, the density of states $\frac{dE}{dn}$ in
(\ref{inverse}) can be computed by means of the inverse function derivative 
rule, i.e. $\frac{dE}{dn} = 1/\frac{d\pi}{dE}$. To proceed further in the
explicit computation of $W(x)$, let us briefly recall some properties of the 
function $\pi(E)$ (see, for instance \cite{Zagier}). Its first estimate 
was obtained by Gauss and Legendre \cite{Legendre}
\EQ
\pi(E) \sim \frac{E}{\ln E} \,\,\,.
\label{Gauss1}
\EN 
A more precise version of the above estimate is given by  
\EQ
\pi(E) \simeq {\makebox Li}(E) \equiv \int_2^E \frac{dE'}{\ln E'}\,\,\, , 
\label{logarithm}
\EN 
while a further refinement is provided by the series 
\EQ
\pi(E)\, \simeq \,R(E) \, = \, \sum_{m=0}^{\infty} \frac{\mu(m)}{m}
\,{\makebox Li}\left(E^{1/m}\right)  \,\,\, ,
\label{riemann}
\EN 
with the Moebius numbers $\mu(m)$ defined by 
\[
\mu(m) \,=\,
\left\{
\begin{array}{ll}
1 & \mbox {if $ m =1 $} \\
0 & \mbox {if $m$ is divisible by a square of a prime} \\
(-1)^k & \mbox{otherwise}
\end{array}
\right.
\]
where $k$ is the number of prime divisors of $m$. $R(E)$ is the smooth 
function which approximates $\pi(E)$ more efficiently (see, e.g.
\cite{Zagier}). For its derivative we have  
\EQ
\frac{d\pi}{dE} \simeq 
\frac{1}{\ln E} \,
\sum_{m=1}^{\infty} \frac{\mu(m)}{m} \,E^{\frac{1-m}{m}} \,\,\,.
\label{derivative}
\EN 
Hence, in the semiclassical approximation the potential whose energy 
eigenvalues are prime numbers is expressed in terms of the series  
\EQ
x(W)\,=\,
\frac{\hbar}{\sqrt{2 m}} \sum_{m=1}^{\infty} \frac{ \mu(m)}{m}\,
\int_{E_0}^W \,\frac{ {\cal E}^{\frac{1-m}{m}} }{\ln {\cal E} \,
\sqrt{W - {\cal E}} }\,\,d{\cal E} \,\,\, . 
\label{solution}
\EN 
The plot of this function is drawn in Fig.\,1 (with $E_0 = 0$): the 
series (\ref{solution}) rapidly converges to a limiting function, which 
can be regarded as the potential $W(x)$, solution of the problem\footnote{The
degree of accuracy which can be reached in the determination of $W(x)$ is 
purely a question of practical considerations.}. 

The existence of a potential which admits all the prime numbers as 
its only eigenvalues has some important implications. For instance, 
a long--standing issue in number theory such as the primality test of a 
given number $N$ can be completely resolved by the sole application of 
physical laws. To do this, it would be sufficient to design an apparatus 
$G$, which provides the potential shown in Fig.\,2, where the central 
part is nothing else but the potential $W(x)$, truncated however at an 
energy cutoff $\epsilon_0$ that can be controlled by an external handle 
(see Fig.\,3). If the barriers $B$ are shaped such that their penetration 
is strongly inhibited, the original energy levels of $W(x)$ are essentially
left unperturbed: the device $G$ then allows the typical resonance
phenomena of quantum mechanics to be observed (see, e.g. \cite{merzbacher}). 
Hence, to determine whether a given number $N$ is prime or not, we merely 
have to send an incident plane wave of energy $E = N \hbar \omega$ into 
the apparatus $G$: if the number $N$ is a prime number (i.e. if the energy 
$E$ belongs to the spectrum of the potential $W(x)$), then a sharp 
resonance peak in the transmission amplitude $T(E)$ is observed, i.e. the
plane wave will be in practice completly transmitted, otherwise it will 
be simply reflected. For the success of this experiment, we have only 
to pay attention to the tuning of the threshold $\epsilon_0$: this 
has to be choosen so that the resonance peaks of the tunnelling effects 
remain visible but, at the same time, without affecting the energy
eigenvalues which are lower or close to the value $E= N \hbar \omega$ 
under the probe. This condition can always be satisfied by tuning 
$\epsilon_0$ to be sufficiently larger than the input energy $E$. 

\vspace{15mm} {\em Acknowledgements}. I would like to thank G. Ferretti, 
G. Ghirardi, E. Tosatti and R. Zecchina for conversations and comments 
about this {\em divertissement}. I am particularly grateful to 
S. Baroni for helpful discussions on the role of the cutoff $\epsilon_0$ 
in the setting--up of resonance experiments.

\newpage

\hs

{\bf Figure Caption}

\vspace{5mm}

\begin{description}
\item [Figure 1]. Profiles of $W(x)$ (in units of $\hbar \omega$) versus 
$x$ (in units of $\sqrt{\hbar / 2 m \omega}$), obtained by including 
100 terms of the series (\ref{solution}).  
\item [Figure 2]. The potential $V(x)$ as realized by the apparatus G, 
i.e. the prime--number potential $W(x)$ with an energy cut--off 
$\epsilon_0$ and external barriers. 
\item [Figure 3]. Resonance experiment to implement the primality testing:
the device $G$ gives the transmission amplitude $T(E)$ of the incident 
plane wave as an output signal. 
\end{description}

\pagestyle{empty}
\newpage
\begin{figure}
\null
\centerline{
\psfig{figure=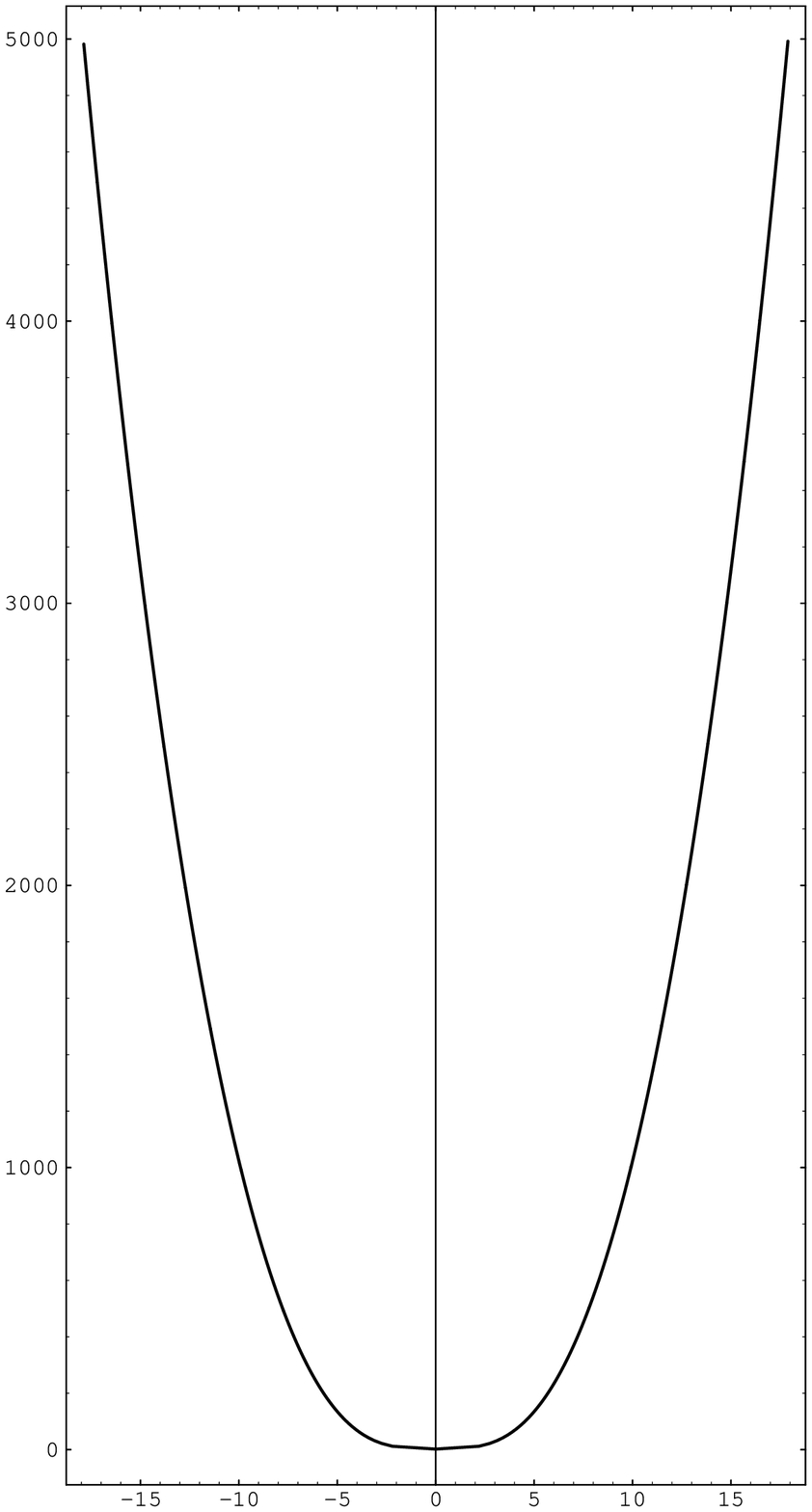}}
\vspace{1cm}
\begin{center}
{\bf \large{Figure 1}}
\end{center}
\end{figure}

\pagestyle{empty}
\newpage
\begin{figure}
\null\vskip 35mm 
\centerline{
\psfig{figure=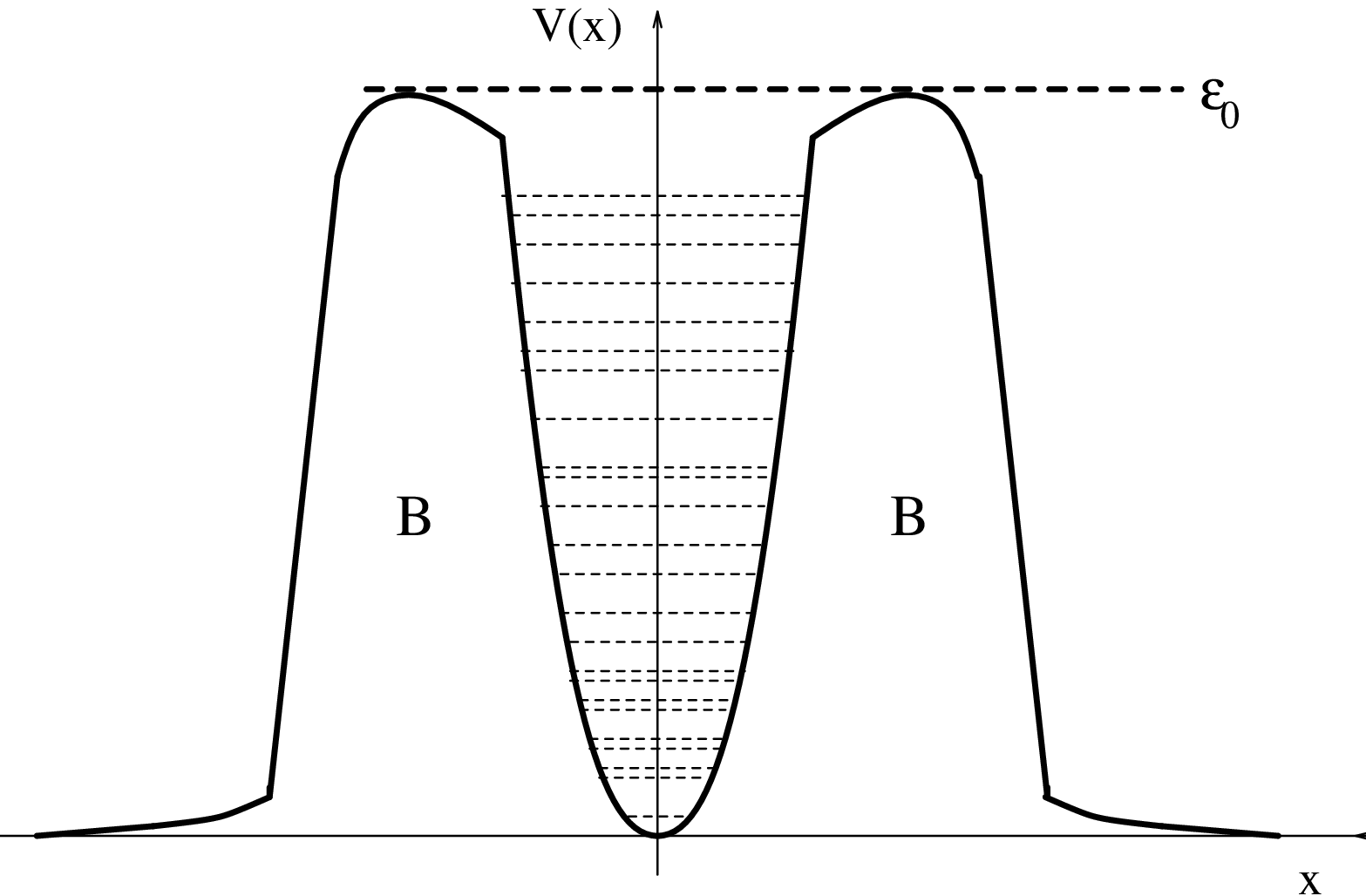}}
\vspace{1cm}
\begin{center}
{\bf \large{Figure 2}}
\end{center}
\end{figure}

\pagestyle{empty}
\newpage
\begin{figure}
\null\vskip 35mm 
\centerline{
\psfig{figure=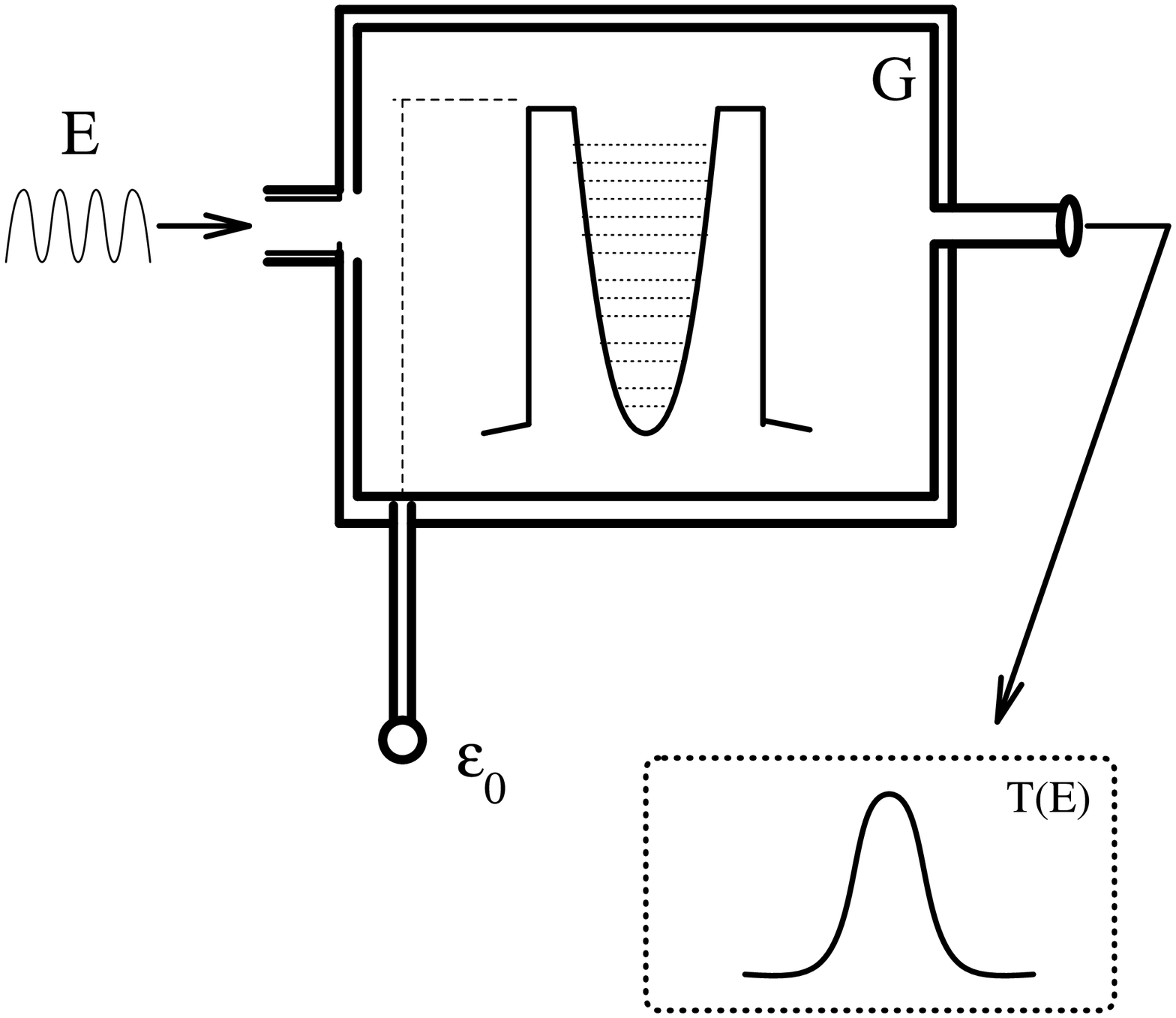}}
\vspace{1cm}
\begin{center}
{\bf \large{Figure 3}}
\end{center}
\end{figure}


\begin{thebibliography}{99}
\bibitem{polygonal} The polygonal numbers $P_s^n$ count the number 
of balls that can be arranged in a regular polygonal pattern of $s$ 
side of increasing rank $n$. Their closed formula is given by 
$p_s^n = \frac{n}{2}\left[ (n-1) s - 2(n-2)\right]$.
\bibitem{perfect} A perfect number is one which is equal to the sum 
of its aliquot divisors. A close formula for even perfect number is due 
to Euclid, ${\cal P}_n = 2^{n-1} \,{\cal M}_n$ where $n$ may be any positive 
integer exceeding unity which makes the second factor ${\cal M}_n 
\equiv (2^n-1)$ a prime number. Integers of the form ${\cal M}_n = (2^n
-1)$ are called Mersenne's numbers and they provide the largest prime 
numbers known so far. No odd perfect numbers have been presently found.
\bibitem{Fibonacci} Fibonacci numbers are defined by the recursive 
equation $F_{n+2} = F_{n+1} + F_{n}$, with initial condition $F_1 = 1$, 
$F_2 = 1$. Their closed expression is provided by Binet's formula 
$F_n = \frac{1}{\sqrt 5}
\left[\left(\frac{1+\sqrt{5}}{2}\right)^n 
- \left(\frac{1-\sqrt{5}}{2}\right)^n \right]$.
\bibitem{Legendre} Although formula (\ref{Gauss1}) is 
just a coarse approximation, it is known however that it is able 
to capture the asymptotic behaviour of $\pi(E)$ -- a result which
constitutes the content of the ``Prime Number Theorem'' 
\cite{primenumbertheorem}  
\[
\lim_{E\to\infty} \frac{\pi(E) \ln E}{E} \,=\,1 \,\,\,.
\label{theorem}
\]
\end{thebibliography}
\end{document}